\documentclass[%
reprint,
showpacs,preprintnumbers,
 amsmath,amssymb,
 aps,
 pra,
floatfix,
]{revtex4-1}
\usepackage{color}
\usepackage{graphicx}
\usepackage{dcolumn}
\usepackage{bm}

\newcommand{\bra}[1]{\left\langle #1 \right\vert}
\newcommand{\ket}[1]{\left\vert #1 \right\rangle}
\newcommand{\dotprod}[2]{\left\langle #1 \middle| #2 \right\rangle}
\newcommand{\expect}[3]{\left\langle #1 \middle| #2 \middle| #3 \right\rangle}
 
\begin{document}
\title{Adiabatic electronic flux density: a Born-Oppenheimer Broken Symmetry ansatz}

\author{Vincent Pohl}
 \email{v.pohl@fu-berlin.de}
\author{Jean Christophe Tremblay}
\affiliation{%
 Institute for Chemistry and Biochemistry, Freie Universit\"at Berlin, Takustra\ss e 3, 14195 Berlin, Germany
}%

\date{\today}

\begin{abstract}
The Born-Oppenheimer approximation leads to the counterintuitive result 
of a vanishing electronic flux density upon vibrational dynamics in the electronic ground state.
To circumvent this long known issue, we propose using pairwise anti-symmetrically translated vibronic densities
to generate a symmetric electronic density that can be forced to satisfy the continuity equation approximately.
The so-called Born-Oppenheimer broken symmetry ansatz yields all components of the flux density simultaneously
while requiring only knowledge about the nuclear quantum dynamics on the electronic adiabatic ground state potential energy surface.
The underlying minimization procedure is transparent and computationally inexpensive, and
the solution can be computed from the standard output of any quantum chemistry program.
Taylor series expansion reveals that the implicit electron dynamics originates from non-adiabatic coupling to the explicit
Born-Oppenheimer nuclear dynamics.
The new approach is applied to the ${\rm H}_2^+$ molecular ion vibrating in its ${}^2\Sigma^+_g$ ground state.
The electronic flux density is found to have the correct nodal structure and symmetry properties at all times.

\end{abstract}

                              
\maketitle
 
\section{Introduction}

Within the Born-Oppenheimer framework \cite{27:BO:boa}, 
molecules are described as moving adiabatically on a potential energy landscape
defined by the system electrons \cite{35:E:tst,35:EP:tst,87:P:pes}.
Quantitative knowledge of the electronic flux during molecular processes could lead to a deeper understanding of their underlying mechanisms.
Interpreting the electronic density as a probability fluid, the flux of electrons is usually described as the amount of probability crossing
dividing surfaces per unit time \cite{01:M:ED_as_fluid}. 
As such, the electronic flux is defined from a scalar field, the electron flow, that can be reconstructed in principle from the time evolution of 
an experimentally observed electronic probability 
\cite{12:DVS:ED_exp,13:DSMS:ED_exp,14:SK:ED_exp}. 
This was already achieved experimentally for the nuclear density analogon \cite{00:FB:Na2_Flux_exp,06:ERFZSMU:D2+_Flux_exp,13:MJY:NFD_exp}.
From a theoretical perspective, the determination of electronic fluxes suffers from the requirement of defining dividing surfaces in 
nuclear configuration space, for which the definition of a partitioning scheme is not unique. This problem is also encountered in, e.g.,
the determination of individual atomic charges in molecules \cite{04:GHBB:VDD}, and it is exacerbated by the fact that nuclei are moving during the molecular
process of interest and that their position is only defined as a quantum mechanical distribution, as opposed to a single point in nuclear configuration space.

To complement the information obtained from the electronic fluxes, the knowledge of the time-dependent electronic flux densities can significantly 
improve understanding of these processes at a microscopic level. 
The latter object corresponds to a vector field describing the instantaneous displacement of probability fluid elements at every point in electronic
configuration space, and it does not rely on a particular nuclear spatial partitioning scheme.
In conventional quantum molecular dynamics simulations, electrons are adiabatically separated from the motion of the nuclei because 
of their large mass mismatch. The solution to the Schr\"odinger equation leads to a real-valued, stationary electronic wavefunction for each given
nuclear configuration. This has the unfortunate consequence of leading to a vanishing electronic flux density \cite{26:Schroedinger}. A workaround to
this unphysical result can be obtained by means of a vibronic Born-Huang expansion \cite{bornhuang}, where the nuclear dynamics is performed on multiple
coupled potential energy surfaces simultaneously.
Since the determination of the electronic flux density beyond the Born-Oppenheimer approximation are computationally demanding,
only very few molecular systems, H$_2^+$\cite{05:P:NBO_H2+,06:KSKNK:NBO_H2+,06:KH:NBO_H2+,12:NTM:NBO_H2+,12:INN:NBO_H2+,13:Jhon:NBO_H2+}, 
H$_2$ \cite{09:BLSA:NBO_H2_GS,95:CA:NBO_H2}, and H$_2$D$^+$\cite{11:MHMR:NBO_H2D+,12:MR:NBO_H2D+}, have been investigated up to date. Hence,
new approximate schemes are required to overcome this problem.

A point of particular interest is that electronic flux densities computed using the Born-Huang ansatz reveal that the nuclear dynamics
from which it emerges
almost quantitatively follows the ground state molecular dynamics for small amplitude vibrations, contrary to the adiabatic Born-Oppenheimer picture.
To circumvent this long known issue, a few workarounds have been proposed that suffer from various
drawbacks. In the semi-classical coupled channels theory \cite{12:D:CCa,12:DKMP:CCb,13:DKMPPP:SCC,14:HPPP:H2_EFD}, the electronic flux density strictly follows 
the nuclear motion, while the time-shift classical approach of Okayama and Takatsuka \cite{09:OT:time_shift_flux} yields a complex-valued
flux density.
Other attempts at reducing the complete Schr\"odinger equation provide information about
individual components of the flux density in the average field of the others \cite{15:Manz_etal:review_1d_EFD}.
Perturbative approaches including the effect of multiple excited electronic states have also been
put forward with various degree of success \cite{83:N:EFD_perturb,12:P:EFD_perturb,13:D:EFD_perturb}, but their framework departs from the Born-Oppenheimer picture.

In an attempt to reconcile the adiabatic approximation with the intuitive picture
of electrons flowing along with the nuclear motion, we present
an alternative ansatz correlating the electronic with the nuclear motion.
Starting from the Liouville von Neumann equation for the evolution of the vibronic density matrix operator,
we first reveal inconsistencies in the Born-Oppenheimer approximation.
In order to rectify these, pairwise antisymmetric translation operators are introduced to shift the density matrix operator within the nuclear
configuration space. 
Tracing out the nuclear degrees of freedom yields an electronic continuity equation stemming from a single reduced electronic density
depending on a single parameter, the so-called correlation length.
The continuity equation can thus be seen satisfied approximately by numerical optimization of a cost equation,
for which a robust and computationally inexpensive implementation is obtained by means of Taylor series expansion.
The method is applied to a first model system, the hydrogen molecular ion ${\rm H}_2^+$ vibrating in the electronic ground state ${}^2\Sigma^+_g$,
for which non Born-Oppenheimer results are available \cite{13:Jhon:NBO_H2+,13:DKMPPP:SCC}.

\section{Theory}

The object of our investigations, the electronic probability density, can be understood as a diagonal element of the reduced
electron density matrix in the position representation. The latter can be obtained from the reduction of the total 
density operator, $\hat{\Theta}(t)$, which evolves according to the Liouville von Neumann equation
\begin{equation}\label{lvn}
  \frac{\partial}{\partial t}\hat{\Theta}(t)=-\frac{\imath}{\hbar}\left[\hat{H}_{\rm el}+\hat{T}_{\rm nuc},\hat{\Theta}(t)\right]
\end{equation}
with the electronic Hamiltonian $\hat{H}_{\rm el}$ and the nuclear kinetic energy operator $\hat{T}_{\rm nuc}$.
For a coherent vibronic system in a pure state, this density matrix operator  $\hat{\Theta}(t)$ can be factorized as
\begin{equation}
  \hat{\Theta}(t) = \ket{\psi(t)}\bra{\psi(t)}.
\end{equation}
Within the Born-Oppenheimer approximation, the system wavefunction $\ket{\psi(t)}$ is written as
\begin{equation}\label{boa}
  \ket{\psi(t)} = \ket{\varphi}\ket{\chi(t)},
\end{equation}
with the time-dependent nuclear wavefunction $\ket{\chi(t)}$ and the time-independent electronic wavefunction $\ket{\varphi}$.
The latter is defined by the time-independent electronic Sch\"odinger Equation 
$\hat{H}_{\rm el} \ket{\varphi} = V_{\rm tot} \ket{\varphi}$. Here, the electronic Hamiltonian 
$\hat{H}_{\rm el} = \hat{T}_{\rm el}+\hat{V}_{\rm eff}$
refers to an effective one-electron operator in the spirit of the density functional theory, i.e., we define an
effective potential energy operator $\hat{V}_{\rm eff}$ as a time-dependent multiplicative potential according to the
Runge-Gross theorem \cite{84:RG:tddft}.

The electronic probability density at an observation point $r$ can be defined by the trace
over the nuclear contributions, here given in the position representation
\begin{eqnarray}
  \rho_{\rm el}(r,t) & = & \int {\rm d}Q \bra{r,Q}\hat{\Theta}(t)\ket{r,Q}\label{rhoel_1}\\
  & = & \int {\rm d}Q  \dotprod{r,Q}{\psi(t)}\dotprod{\psi(t)}{r,Q}\nonumber\\
  & = & \int {\rm d}Q\, \vert\varphi(r;Q)\vert^2\vert\chi(Q,t)\vert^2.\label{rhoel_2}
\end{eqnarray}
Note that the nuclear wavefunction depends on time $t$ and the nuclear coordinate $Q$, and
the electronic wavefunction depends on the electronic coordinate $r$ and parametrically
on the nuclear coordinate $Q$. 

\subsection{Time evolution of the electron density within the Born-Oppenheimer approximation}\label{boa_eom}

To reveal a fundamental problem of the Born-Oppenheimer approximation, expressions for the time evolution of the electron density
are deduced following two different routes: taking the time-derivative of the electronic density after reduction of the total density matrix
[cf. Eq. \eqref{rhoel_2}], or reducing the equations of motion after the applying the respective operators to the wavefunctions
[cf. Eqs. \eqref{lvn} and \eqref{rhoel_1}].
For clarity, only a system consisting of one effective nuclear coordinate $Q$ with mass $\mu_Q$
is considered throughout. The procedure can be easily extended to a higher dimensionality.

Based on Eq. (\ref{rhoel_2}), the time evolution of the electronic density can be written as
\begin{equation}\begin{aligned}\label{dtrhoel_1}
  \frac{\partial}{\partial t}\rho_{\rm el}(r,t) &= \frac{\partial}{\partial t} \int {\rm d}Q \vert\varphi(r;Q)\vert^2\vert\chi(Q,t)\vert^2\\
  &= \int {\rm d}Q \vert\varphi(r;Q)\vert^2\frac{\partial}{\partial t}\Big(\vert\chi(Q,t)\vert^2\Big).\\  
\end{aligned}\end{equation}
Using the time-dependent nuclear Sch\"odinger equation,
\begin{equation}\begin{aligned}\label{TDSE_nuc}
  \frac{\partial}{\partial t} \chi(Q,t) & = -\frac{\imath}{\hbar} \big(\hat{T}_{\rm nuc}+V_{\rm tot}\big) \chi(Q,t),
\end{aligned}\end{equation}
the time derivative of the nuclear density leads to the nuclear continuity equation
\begin{equation}\begin{aligned}
  \frac{\partial}{\partial t} \vert\chi(Q,t)\vert^2 & = \chi(Q,t) \frac{\partial}{\partial t} \chi^\dagger(Q,t) + \chi^{\dagger}(Q,t) \frac{\partial}{\partial t} \chi(Q,t)\\
  & = \frac{\imath\hbar}{2\mu_Q} \left(\chi^\dagger(Q,t)\nabla_Q^2\chi(Q,t)-{\rm c.c.}\right)  \\ 
  & = -\nabla_Q \cdot \vec{j}_{\rm nuc},
\end{aligned}\end{equation}
where ``c.c.`` stands for the complex conjugate and $\vec{j}_{\rm nuc}$  is the nuclear flux density.
Inserting the result into Eq. \eqref{dtrhoel_1} yields
\begin{equation}\begin{aligned}\label{dtrhoel_1_end}
  \frac{\partial}{\partial t}\rho_{\rm el}(r,t) &= - \int {\rm d}Q  \Big(\nabla_Q \vert\varphi(r;Q)\vert^2\Big) \cdot \vec{j}_{\rm nuc}. 
\end{aligned}\end{equation}
In deriving Eq. \eqref{dtrhoel_1_end}, the divergence theorem was applied to eliminate a vanishing contribution from the bound vibrational states.
This electronic flow cannot be trivially expressed as the divergence of 
an electronic vector field, which is a requirement for formulating an expression for an {\it electronic flux density}.

As an alternative route starting from Eq. \eqref{rhoel_1}, the time evolution of the density matrix operator $\hat{\Theta}(t)$ 
is represented in the position representation using Eq. \eqref{lvn}
\begin{widetext}
\begin{equation}\begin{aligned}\label{dtrhoel_2}
  \frac{\partial}{\partial t}&\rho_{\rm el}(r,t) = \int {\rm d}Q  \expect{r,Q}{\frac{\partial}{\partial t} \hat{\Theta}(t)}{r,Q} \\
  = & -\frac{\imath}{\hbar} \int {\rm d}Q \Big(\bra{r,Q}\hat{T}_{\rm el}\hat{\Theta}(t)\ket{r,Q}-\bra{r,Q}\hat{\Theta}(t)\hat{T}_{\rm el}\ket{r,Q}
  +\bra{r,Q}\hat{T}_{\rm nuc}\hat{\Theta}(t)\ket{r,Q}-\bra{r,Q}\hat{\Theta}(t)\hat{T}_{\rm nuc}\ket{r,Q}\\
  &+\bra{r,Q}\hat{V}_{\rm eff}\hat{\Theta}(t)\ket{r,Q}-\bra{r,Q}\hat{\Theta}(t)\hat{V}_{\rm eff}\ket{r,Q}\Big).
\end{aligned}\end{equation}
\end{widetext}
Here, the last line of Eq. (\ref{dtrhoel_2}) vanishes in the position representation, since $\hat{V}_{\rm eff}$ is a multiplicative operator.
At this stage, the electronic density still obeys both the Liouville von Neumann and the total continuity equations, as no assumption on the dynamics nor 
the specific form of the density operator has been made.
Utilizing the Born-Oppenheimer ansatz [cf. Eq. \eqref{boa}] for a density matrix representing a system in a pure state yields a vanishing electron flow
\begin{widetext}
  \begin{equation}\begin{aligned}\label{dtrhoel_2_boa}
    \frac{\partial}{\partial t}&\rho_{\rm el}(r,t) = -\frac{\imath}{\hbar} \int {\rm d}Q \vert\chi(Q,t)\vert^2 \left\lbrace \varphi(r;Q)\left(-\frac{\hbar^2}{2m_e}\vec{\nabla}_{e}^{2}\varphi(r;Q)\right)
    - \varphi(r;Q)\left(-\frac{\hbar^2}{2m_e}\vec{\nabla}_{e}^{2}\varphi(r;Q)\right) \right\rbrace \\  
    & -\frac{\imath}{\hbar} \int {\rm d}Q \left\lbrace \varphi(r;Q)\chi^\dagger(Q,t)\left(-\frac{\hbar^2}{2\mu_Q}\nabla_{Q}^{2}\right)\Big(\varphi(r;Q)\chi(Q,t)\Big) 
    - \varphi(r;Q)\chi(Q,t)\left(-\frac{\hbar^2}{2\mu_Q}\nabla_{Q}^{2}\right)\Big(\varphi(r;Q)\chi^\dagger(Q,t)\Big)\right\rbrace \\  
    =&~ \frac{\imath\hbar}{2\mu_Q} \int {\rm d}Q \,\nabla_Q \cdot \bigg( \varphi(r;Q)\chi^\dagger(Q,t)\nabla_Q\Big(\varphi(r;Q)\chi(Q,t)\Big) - {\rm c.c.}\bigg)\\
    =&~ 0,
  \end{aligned}\end{equation}  
\end{widetext}
where $\vec{\nabla}_e$ denotes the gradient with respect to the electronic coordinates $r$ and $m_e$ refers to the electronic mass.
The last line of Eq. \eqref{dtrhoel_2_boa} arises from the divergence theorem and is valid for bound vibrational states. 
Eq. \eqref{dtrhoel_1_end} stands in strong contradiction to Eq. \eqref{dtrhoel_2_boa}: although the time evolution of the {\it a priori} reduced
electron density shows a non-vanishing flow, there is neither a flow, nor a flux density in Eq. \eqref{dtrhoel_2_boa}. 
This implies that the reduced electron density in Eq. \eqref{dtrhoel_1_end} does not implicitly satisfy the Liouville von Neumann equation. 
This is because the {\it a priori} reduced electron density matrix is real and does not represent a pure state density matrix anymore.

\subsection{Breaking the symmetry in the equation of motions}\label{bs}

In order to fix this contradiction, which was already recognized by others (cf. for example
\cite{12:D:CCa,12:DKMP:CCb,13:DKMPPP:SCC,09:OT:time_shift_flux,:M:EFD_Manz,83:N:EFD_perturb,12:P:EFD_perturb,13:D:EFD_perturb}), 
we propose to break the symmetry of the equations of motion.
To this end, a translation operator with the following properties is introduced
\begin{equation}\begin{aligned}\label{unitary}
  U_Q\ket{Q} = \ket{Q + \delta_q}~;&~U_Q^{\dagger}\ket{Q} = \ket{Q - \delta_q}\\
  U_Q^{\dagger}~U_Q=&~U_Q~U_Q^{\dagger} = 1\\
\end{aligned}\end{equation}
and a correlated electron density is defined
\begin{equation}\label{rho_c}
  \rho_{c} = \frac{1}{2}\left(\rho_{+}+\rho_{-}\right).
\end{equation}
To simplify the notation, the coordinate dependence in all quantities is omitted, i.e., 
$\rho_{+}(r,t)\equiv\rho_{+}$, $\varphi(r;Q+\delta_q)\equiv\varphi^{+\delta_{q}}$, and $\chi(Q+\delta_q,t)\equiv\chi^{+\delta_{q}}$. 
Here, a symmetric electron density $\rho_{c}$ is recovered by summing over the pairwise antisymmetric densities $\lbrace\rho_{+},~\rho_{-}\rbrace$.
\begin{figure}[tb]
\includegraphics[width=0.8\linewidth]{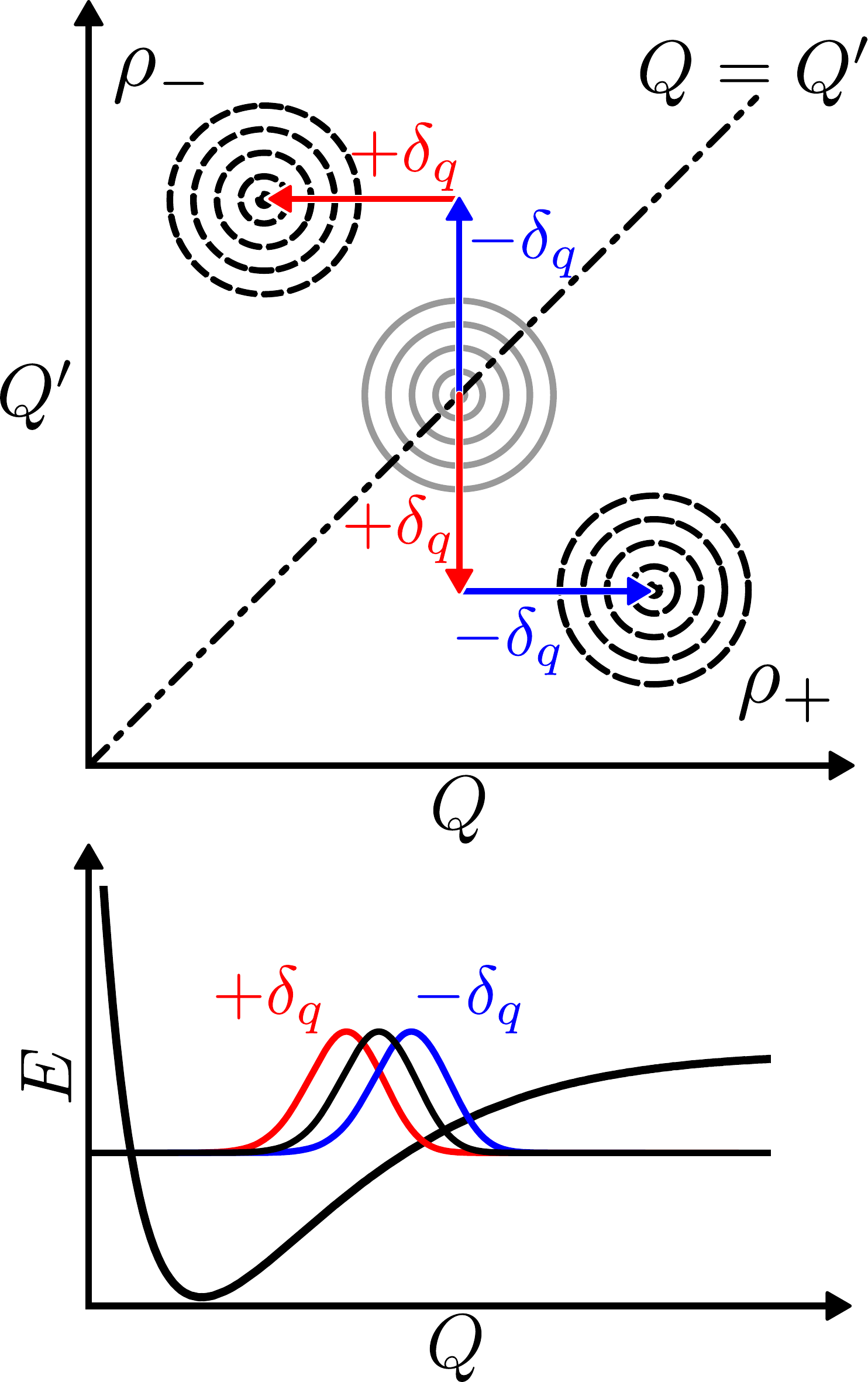}
\caption{\label{fig:concept} (Color online)
Schematic representation of the Born-Oppenheimer broken symmetry ansatz. The bottom panel shows the two translated vibrational wavefunctions
used to generate the broken symmetry densities. Top panel: a symmetric vibronic density is created by adding two pairwise antisymmetric densities
with respect to the $Q=Q'$ axis. The coordinate $Q'$ is introduced to emphasize that two different translation are used in Eq. \eqref{rho_c}.
}
\end{figure}
Positive displacement of the density along the nuclear coordinate $Q$ yields in the Born-Oppenheimer approximation
\begin{equation}\begin{aligned}    
  \rho_{+} & =\int {\rm d}Q \dotprod{r,Q}{U_{Q}\hat{\Theta}(t) U_{Q}\middle\vert r,Q} \\
    & =\int {\rm d}Q \dotprod{r,Q-\delta_q}{\hat{\Theta}(t)\middle\vert r,Q+\delta_q}\\
  & =\int {\rm d}Q \left(\varphi^{-\delta_{q}}\varphi^{+\delta_{q}}\right)\left(\chi^{-\delta_{q}}\left(\chi^{+\delta_{q}}\right)^{\dagger}\right)\\
  & =\int {\rm d}Q \left(\varphi^{-\delta_{q}}\chi^{-\delta_{q}}\right)\left(\varphi^{+\delta_{q}}\chi^{+\delta_{q}}\right)^{\dagger}, \label{displaced_boa}
\end{aligned}\end{equation}
and negative displacement results in $\rho_{-}$, which can be constructed analogously.
The physical meaning given to $\delta_q$ is that of electron-nucleus correlation
in nuclear configuration space, which we dub {\it correlation length}. 
Note that the transformation $U_{Q}\hat{\Theta}(t)U_{Q}$ is not unitary.
The cartoon in the top panel of Fig. \ref{fig:concept} illustrates how two pairwise antisymmetric densities created from the translation of vibronic
wavefunctions at positive and negative correlation lengths $\delta_q$ can be used to generate a symmetric density.
The vibrational part of the wavefunction and its translations are depicted in the bottom panel. 
The coordinate $Q'$ in the top panel has been introduced to emphasize the fact that the density is computed as a
product of two translated wavefunctions in opposite directions.
By translating the density matrix $\langle r,Q'\vert \hat{\Theta}(t)\vert r,Q \rangle$,
spatial correlation within the nuclear configuration space is 
implicitly transfered from the off-diagonal elements of the vibronic density matrix
to the diagonal elements of the reduced electron density matrix.

\subsection{The correlated electronic continuity equation}\label{nonlinearBOBS}

Provided the broken symmetry reduced electronic density implicitly satisfies the Liouville von Neumann equation for the total density matrix,
Eq. \eqref{lvn}, the definition of an electron flow should be independent of the order in which the reduction procedure is performed
[see Section \ref{boa_eom}]. That is, substituting the correlated density for the electronic density in Eq. \eqref{rhoel_2}
and tracing out the nuclear degrees of freedom yields the following flow components
\begin{equation}\begin{aligned}\label{jc_left}
  \frac{\partial}{\partial t} \rho_{c} &=\frac{1}{2}\left(\frac{\partial}{\partial t} \rho_{+} + \frac{\partial}{\partial t} \rho_{-}\right)\\
  \frac{\partial}{\partial t} \rho_{\pm} &= \int {\rm d}Q \left(\varphi^{\mp\delta_{q}}\varphi^{\pm\delta_{q}}\right)\frac{\partial}{\partial t}\left(\chi^{\mp\delta_{q}}\left(\chi^{\pm\delta_{q}}\right)^{\dagger}\right)
\end{aligned}\end{equation}
Following the alternative route using the Liouville von Neumann equation as described in Eq. (\ref{dtrhoel_2}) and Eq. (\ref{dtrhoel_2_boa}) 
with the correlated density yields a non-vanishing flow with components
\begin{equation}\begin{aligned}\label{jc_right}
  \frac{\partial}{\partial t} \rho_{\pm} &=  -\frac{\imath\hbar}{2m_{e}}\vec{\nabla}_{e}\cdot\int {\rm d}Q \\
  &~ \times \left(\varphi^{\mp\delta_{q}}\vec{\nabla}_{e}\varphi^{\pm\delta_{q}}-\varphi^{\pm\delta_{q}}\vec{\nabla}_{e}\varphi^{\mp\delta_{q}}\right)\\
   &~ \times \left(\chi^{\mp\delta_{q}}\left(\chi^{\pm\delta_{q}}\right)^{\dagger}\right).
\end{aligned}\end{equation}
Note that two components appear naturally as the divergence of an electronic vector field which allows for a unique definition of the flux density
as the sum of two broken symmetry terms
\begin{equation}\begin{aligned}
\vec{j}_{\pm} &= \frac{\imath\hbar}{2m_{e}}\int {\rm d}Q\left(\varphi^{\mp\delta_{q}}\vec{\nabla}_{e}\varphi^{\pm\delta_{q}}-\varphi^{\pm\delta_{q}}\vec{\nabla}_{e}\varphi^{\mp\delta_{q}}\right) 
  \\
  &\times\left(\chi^{\mp\delta_{q}}\left(\chi^{\pm\delta_{q}}\right)^{\dagger}\right)~.
\end{aligned}\end{equation}
Combining Eqs. \eqref{jc_left} and \eqref{jc_right} yields the following correlated electronic continuity equation
\begin{equation}\begin{aligned}\label{jc_ceq}
  \frac{\partial}{\partial t} \rho_{c} &=  -\vec{\nabla}_{e}\cdot\vec{j}_{c} \\ 
  \frac{1}{2}\left(\frac{\partial}{\partial t} \rho_{+} + \frac{\partial}{\partial t} \rho_{-}\right) &= -\frac{1}{2}\left(\vec{\nabla}_{e}\cdot\vec{j}_{+} 
+\vec{\nabla}_{e}\cdot\vec{j}_{-} \right).\\ 
\end{aligned}\end{equation}
Since the validity of Eq. (\ref{jc_ceq}) depends only on a single control parameter, the correlation length $\delta_{q}$, it may be uniquely determined 
at each timestep by numerical minimization of the residual cost functional
\begin{equation}\label{opt}
  \min_{\delta_{q}\in\Re}\left\|\frac{\partial}{\partial t}\rho_{c}+\vec{\nabla}_{e}\cdot\vec{j}_{c}\right\|_{2}.
\end{equation}
We dub this procedure Born-Oppenheimer Broken Symmetry (BOBS) ansatz.
It can be underlined that within the BOBS ansatz, the same correlated density yields an electron flow on 
the left and an electronic flux density on the right-hand-side of the electronic continuity equation Eq. (\ref{jc_ceq}).

\subsection{Taylor series expansion}\label{linearBOBS}

To reveal the physical origin of the flow and flux density in Eqs. (\ref{jc_left}-\ref{jc_ceq}),
it is instructive to expand all terms using a Taylor series expansion around the point $Q$:
\begin{equation}\begin{aligned}\label{taylor}
  \varphi^{\pm\delta_{q}} & =\varphi\pm\delta_{q}\left.\varphi'\right|_{Q}+\frac{\delta_{q}^{2}}{2}\left.\varphi''\right|_{Q}\pm\dots \\
  \chi^{\pm\delta_{q}} & =\chi\pm\delta_{q}\left.\chi'\right|_{Q}+\frac{\delta_{q}^{2}}{2}\left.\chi''\right|_{Q}\pm\dots
\end{aligned}\end{equation}
where an apostrophe abbreviates a nuclear derivative, e.g., $\chi'=\partial\chi / \partial Q$. 
The series is expected to converge since the correlation length $\delta_q$ is small, as will be seen later.
\begin{figure*}[tb]
\includegraphics[width=0.9\linewidth]{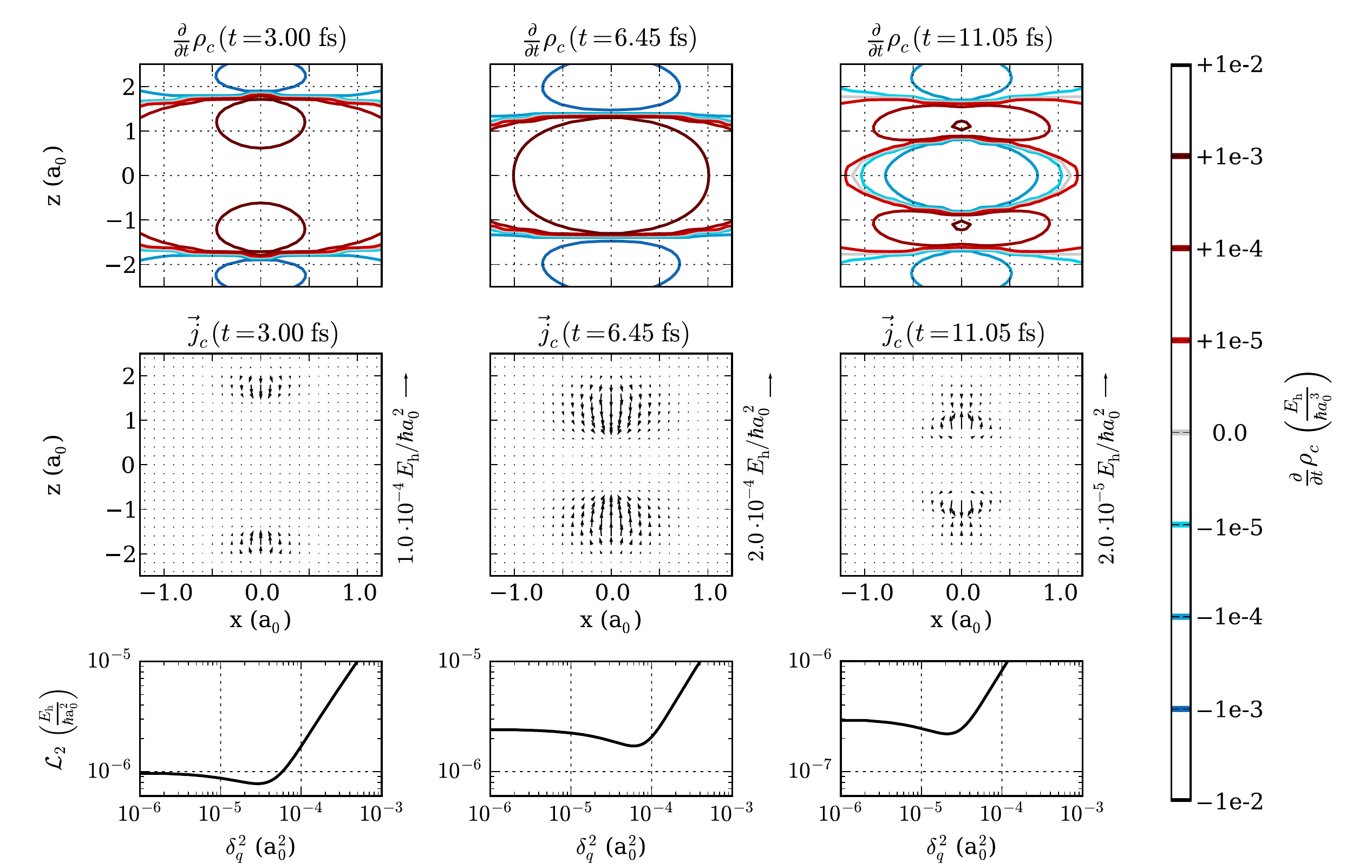}
\caption{\label{fig:snapshots} (Color online)
Representative snapshots of the electron flow (top panel) and the electronic flux density (central panels) computed with the linearized BOBS ansatz [cf. Eqs. (\ref{taylor_jc_left}, \ref{taylor_jc_right})] 
for a vibrating hydrogen molecular ion ${\rm H}_2^+$  oriented along the $z$-axis. 
The results are plotted in the $xz$-plane for three different times within first half vibrational period: 
for $t=3\,{\rm fs}$ (left column), for $t=6.45\,{\rm fs}$ (middle column), and for $t=11.05\,{\rm fs}$ (right column). 
Note the logarithmic scale in the contour plots and the different scales in the vector plots. 
Bottom panels: Error of the correlated electronic continuity equation per grid point, $\mathcal{L}_2=\frac{1}{N}\big\|\frac{\partial}{\partial t}\rho_{c}+\vec{\nabla}_{e}\cdot\vec{j}_{c}\big\|_{2}$, 
as a function of the correlation length squared for the three selected snapshots.
}
\end{figure*}

Substituting Eq. \eqref{taylor} in Eq. \eqref{jc_ceq}, the left and the right-hand-side of the correlated continuity equation take the 
following form to second order 
\begin{equation}\begin{aligned}\label{taylor_jc_left}
  &\frac{\partial}{\partial t}\rho_{c} = \int {\rm d}Q \left(\left|\varphi\right|^{2}+\delta_{q}^{2}\left(\varphi''\varphi-\varphi'\varphi'\right)\right)\frac{\partial}{\partial t}\left(\chi\chi^{\dagger}\right) \\
  &~+\frac{\delta_{q}^{2}}{2}\int {\rm d}Q \left|\varphi\right|^{2}\frac{\partial}{\partial t}\left(\chi''\chi^{\dagger}-2\chi'\chi'^{\dagger}+\chi\chi''^{\dagger}\right)+O\left(\delta_{q}^{4}\right)
\end{aligned}\end{equation}
and
\begin{equation}\begin{aligned}\label{taylor_jc_right}
  -\vec{\nabla}_{e}\cdot\vec{j}_{c} &= -\frac{\imath\hbar\delta_{q}^{2}}{m_{e}}\vec{\nabla}_{e}\cdot\int {\rm d}Q\left(\varphi\vec{\nabla}_{e}\varphi'-\varphi'\vec{\nabla}_{e}\varphi\right)\\
  &\times\left(\chi\chi'^{\dagger}-\chi'\chi^{\dagger}\right)
  +O\left(\delta_{q}^{4}\right).
\end{aligned}\end{equation}
All electronic quantities appearing in the linearized BOBS ansatz remain real-valued and can be obtained from 
standard quantum chemistry programs.
To zeroth order and/or for zero correlation length, the Born-Oppenheimer flow on the left-hand-side remains unchanged and 
the divergence of the flux density vanishes, leading to the aforementioned contradiction.
Higher orders terms lead to a correction to the electron flow and the emergence of a new term on the right-hand-side 
that can be written as the divergence of a vector field.
It is worth noticing that all odd terms in Eq. (\ref{taylor}) vanish due to the symmetrization of the correlated density.
Consequently, a Taylor series expansion to second order 
yields an error of order $O\big(\delta_{q}^{4}\big)$ on both sides of the equation.

The second order corrections to the electron flow, Eq. \eqref{taylor_jc_left}, and to the electronic flux density,  Eq. \eqref{taylor_jc_right},
can be understood as non-adiabatic coupling elements. The first correction to the electron flow is the convolution of the nuclear flow with the nuclear
spread of the electronic wavefunction, $(\varphi''\varphi-\varphi'\varphi')$, which is reminiscent of a Huang term. The second term is 
the convolution of the nuclear spread of the nuclear wavepacket, $(\chi''\chi^{\dagger}-2\chi'\chi'^{\dagger}+\chi\chi''^{\dagger})$,
with the stationary ground state electronic density. The first term contributing to the electronic flux density involves mixed derivatives 
of the electronic and nuclear wavefunctions. Similar terms appear in first order time-dependent perturbation theory treatment of non-adiabatic couplings,
which involved a transfer of momentum from nuclear to electronic degrees of freedom. In Eq. \eqref{taylor_jc_right}, the nuclear derivative of the
ground state electronic wavefunction take place of the usual complex conjugate leading to a non-vanishing electron flow. Note that the antisymmetric 
nuclear term, $(\chi\chi'^{\dagger}-\chi'\chi^{\dagger})$, yields a purely imaginary quantity, so that the total expression is real-valued.

It can be emphasized that, from a numerical perspective, Eqs. \eqref{taylor_jc_left} and \eqref{taylor_jc_right} are easy to handle since
$\delta_q$ only appears as a multiplicative factor to the remaining terms. Nuclear derivatives of the electronic wavefunctions can thus be computed
once prior to the dynamics, which significantly facilitates the determination of the optimal correlation length in Eq. \eqref{opt}.
In the appendix,
equations of motion for the correlated electron density in the basis of vibrational eigenstates of the electronic ground state
are derived. These allow simplifying the evaluation of the nuclear wavefunction derivatives to that of the stationary vibrational eigenstates,
which can be also evaluate once prior to the dynamical simulations.
It is worth noting that, in this basis, the Born-Oppenheimer evolution of the wavepacket is known analytically at all times.
Consequently, the BOBS ansatz is amenable to a simple, robust, and efficient numerical procedure
which is in principle applicable to arbitrarily complex systems.

At this point, the similarities and differences between the BOBS ansatz and the complex ''time-shift`` flux proposed by Okuyama and Takatsuka \cite{09:OT:time_shift_flux} should be compared.
The latter is based on {\it ab-initio} molecular dynamics (AIMD), in which the nuclear positions follow classical trajectories.
The authors break the symmetry of the equations of motion by ''translating`` the electronic wavefunctions against each other in the time domain,
which amounts to a spatial translation in a classical AIMD context. The procedure yields a complex-valued flux density which disappears in the
limit of vanishing time-shift, and the imaginary part is interpreted as an induced  flux density.
This is in stark contrast with the fully quantum mechanical BOBS procedure, where the nuclei are described by a wavefunction and are thus not localized.
Further, since the dynamics is performed in the Born-Oppenheimer framework, a time-shift only affects the phase of the nuclear wavepacket.
Consequently, a time-shifted quantum mechanical ansatz would lead to a vanishing flux density, as demonstrated in Section \ref{boa_eom}.
The spatial translation in the BOBS ansatz involves non-adiabatic coupling terms, and the resulting flow and associated flux density remain real-valued at all times.

\section{Computational details}
As a model system, we consider the hydrogen molecular ion ${\rm H}_2^+$ vibrating in the electronic ground state 
${}^2\Sigma^+_g$ oriented along the $z$-axis. The \textsc{molpro} software \cite{MOLPRO_brief} was used to compute the electronic
ground state wavefunction $\varphi(r;Q)$ at the restricted open-shell Hartree Fock level of theory and using an aug-cc-pVTZ basis on all atoms \cite{H:aug-cc-pVnZ}.
The internuclear distance was scanned in the range $Q\in[0.4,18.06]\,{\rm a}_0$ at an equidistant interval of $\Delta Q=0.02\,{\rm a}_0$.
Using the \textsc{orbkit} software \cite{orbkit}, the ground state wavefunction and its analytical electronic derivatives were evaluated 
in the $xz$-plane, with $x=[-2.2,\,2.2]\,{\rm a_0}$, $z=[-3.5,\,3.5]\,{\rm a_0}$, and $\Delta x =\Delta z =0.1\,{\rm a_0}$.
The nuclear derivatives of the ground state electronic wavefunction were determined numerically.
The nuclear vibrational eigenstates $\chi_n(Q)$ and their derivatives were determined numerically using the \textsc{wavepacket} software \cite{WavePacket}.

In order to initiate a ground state dynamics, we chose the same initial conditions as  J. F. P\'erez-Torres \cite{13:Jhon:NBO_H2+} and 
translated the vibrational ground state wavefunction $\chi_0$ by $+2.0\,{\rm a}_0$ to positive values of $Q$.
The implementation of the correlated electronic continuity equation was achieved in the vibrational eigenstate representation,
as described in Appendix \ref{sec:basis_set}.
The correlation length, $\delta_q$, was determined by minimization the $\mathcal{L}_2$-norm of the electronic continuity equation 
[cf. Eq. (\ref{opt})] in its Taylor series form (linearized BOBS ansatz) with a Newton Conjugate-Gradient algorithm, as implemented in \textsc{scipy} \cite{SciPy}.
Minimization of the cost functional, Eq. \eqref{opt}, based on the non-linear Eqs. \eqref{jc_right} and \eqref{jc_left} yielded almost identical
results as when using the second-order equations of motion and only the latter are reported below as the computationally substantially more efficient alternative.
In order to obtain the shifted nuclear and electronic eigenfunctions for arbitrary values of $\delta_q$, the functions were interpolated along the nuclear
configuration space $Q$ and evaluated the electronic wavefunction for each value of $\delta_q$ on the electronic grid $r$. 
This computationally very demanding task is solely required for the non-linear BOBS ansatz.

\begin{figure}[tb]
\includegraphics{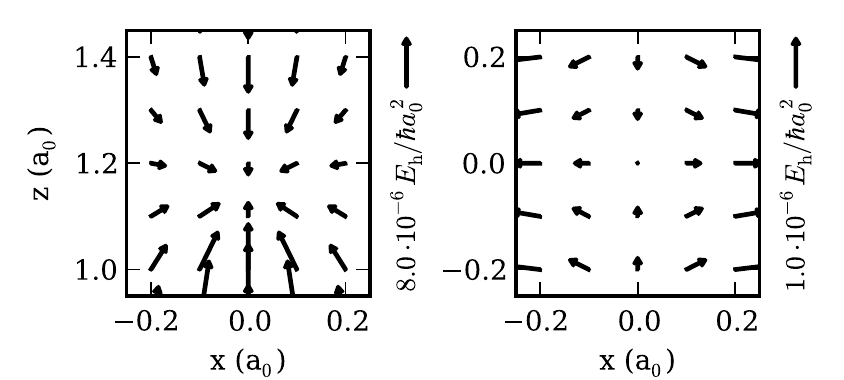}
\caption{\label{fig:nodes}
Detailed view of the symmetry properties of the flux density at the classical turning point, $t=11.05\,{\rm fs}$, at the topmost hydrogen atom (left panel) and 
close to the inversion center (right panel).
The vector field is seen to be antisymmetric with respect to the molecular inversion center and 
the flux density is largest in the pointing towards of the nuclei.
}
\end{figure}

\section{Results and discussion}
\begin{figure}[b]
\includegraphics{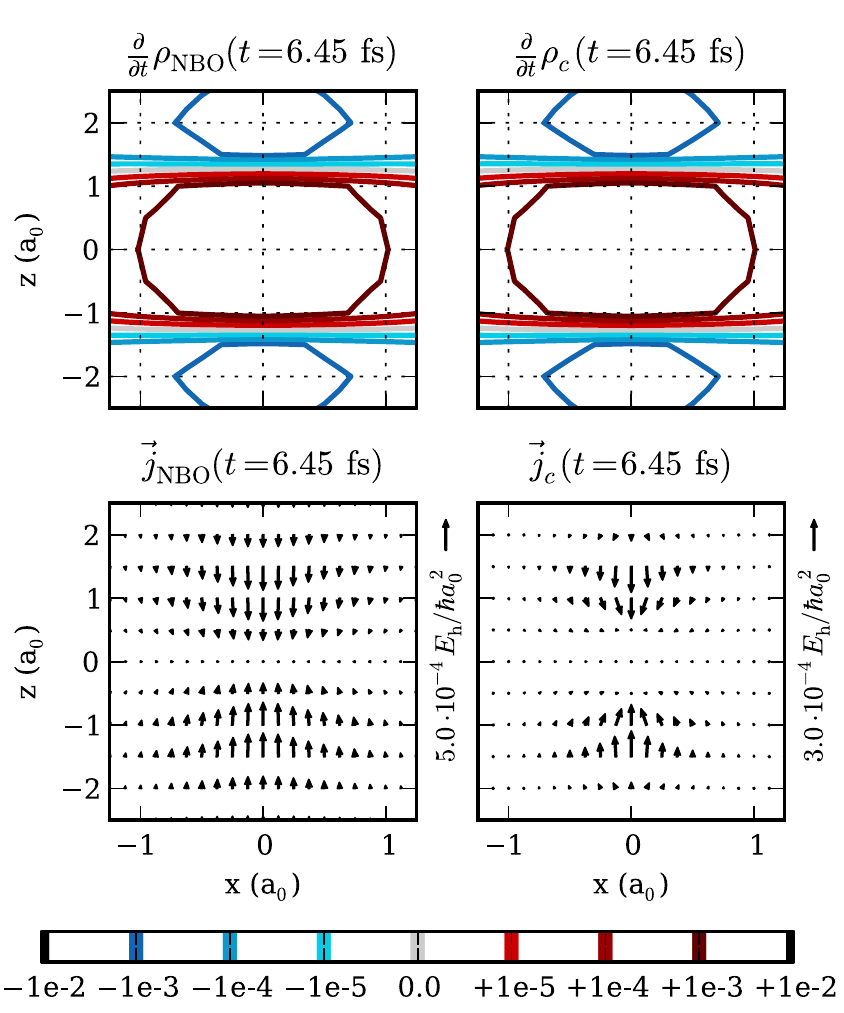}
\caption{\label{fig:JCvsNBO} (Color online)
Representative snapshot at $t=6.45\,{\rm fs}$ of the electron flow (top panels) and the electronic flux density (bottom panels)  in the $xz$-plane
for a vibrating hydrogen molecular ion ${\rm H}_2^+$  oriented along the $z$-axis. 
The results in the left panels are obtained from a non-Born-Oppenheimer ansatz \cite{13:Jhon:NBO_H2+}
and those from the BOBS ansatz are shown in the right panels.
}
\end{figure}

The top panels of Fig. \ref{fig:snapshots} show the time derivative of the correlated electron density, Eqs. (\ref{taylor_jc_left}, \ref{taylor_jc_right}) 
plotted in the $xz$-plane for three representative snapshots within the first half vibrational period. 
For discussion purposes, we define the position of the nuclei as the maximum of the nuclear probability density.
The first two snapshots of the electron flow are smooth and show a nodal 
plane at the nuclei. Those nodal planes are about perpendicular to the vibrational motion before reaching the classical turning point.
During the bond contraction phase (see top left panel, $t=3.00\,{\rm fs}$),
electron density is pulled from behind the nuclei and brought in front to mitigate the increasing nuclear Coulomb repulsion.
This causes a temporary electronic enrichment of the bond, as seen in the top central panels ($t=6.45\,{\rm fs}$). At the turning point of the wavepacket 
(see top right panel, $t=11.05\,{\rm fs}$), the structure of the electron flow becomes more involved because of quantum mechanical interference effects. Density is pulled simultaneously
from the HH bond and behind the atoms towards the nuclei. Consequently, the flow is maximal {\it at} these cusps of the nuclear distribution, 
in contrast to the two other snapshots before reaching
the turning point. Similar observations can be made at longer times, whereas the spread of the wavepacket increases with time and the interference effects become
more pronounced.

The associated flux density obtained from the linearized BOBS ansatz is shown in the central panels of Fig. \ref{fig:snapshots}.
It should be mentioned that the flow shown on the top panels and the divergence of the correlated electronic flux density, Eq. \eqref{taylor_jc_right},
are only in qualitative agreement, with the same nodal structure and maxima of the same magnitude and position. This is a major drawback 
of the BOBS ansatz, for which the continuity equation is only satisfied approximately upon numerical minimization of the cost functional Eq. \eqref{opt}.
On the other hand, the BOBS procedure provides simultaneously an error estimate for the quality of the solution.
At all times, a clear minimum for the cost functional yields a unique definition of the correlation length $\delta_q$.
This is exemplarily illustrated for the three snapshots in the bottom panels of Fig. \ref{fig:snapshots}.
The vector fields of the correlated electronic flux density $\vec{j}_{c}$ (central panels) correlate well with the associated electron flow, pulling the 
density to the bond at early times and towards the nuclei at the turning point.
They also exhibit the correct symmetry properties at all times, i.e., the vector fields are antisymmetric with zero radial flux density $j_{c,x}$ on the
molecular axis (i.e., at $x=y=0$), zero axial flux density $j_{c,z}$ at $z=0$, and no component tangential to the molecular axis.
A detailed view of the electronic flux density symmetry properties in Fig. \ref{fig:nodes} for two different positions demonstrates 
these qualitative features at $t=11.05\,{\rm fs}$.
Furthermore, the flux density is largest in the vicinity of the nuclei, and the axial component is mostly larger than the radial component.
At $t=11.05\,{\rm fs}$, a turning point of the electronic flux density is clearly observable at $z\approx \pm 1.2\,{\rm a_0}$. This is in a good agreement 
with the turning point of the nuclear flux density at $Q\approx 2.4\,{\rm a_0}$. Surprisingly, the electrons are seen to circle around the nuclei
and not exactly moving parallel to the nuclear motion, as observed in previous work \cite{12:D:CCa,12:DKMP:CCb,13:Jhon:NBO_H2+}.

In order to assess the quantitative predictions of the BOBS ansatz, a representative snapshot of the electron flow and the electronic flux density is shown in 
Fig. \ref{fig:JCvsNBO} and compared to the non-Born-Oppenheimer results
for the vibrating hydrogen molecular ion ${\rm H}_2^+$\cite{13:Jhon:NBO_H2+}. From the top panels, it can be recognized that the electron flow computed 
with the BOBS ansatz is only marginally affected by the procedure.
As was shown by others \cite{12:D:CCa,12:DKMP:CCb}, 
this is to be expected since the dynamical evolution of the nuclear wavepacket is properly described within the Born-Oppenheimer approximation.
The BOBS ansatz is a self-consistent procedure based on the numerical optimization
of a parametric, perturbative continuity equation. As such, the results will
be only meaningful if the perturbation introduced by the correlation length remains
small. Provided the nuclear dynamics proceeds on a single Born-Oppenheimer potential energy 
surface, the electron flow is given almost exactly by the evolution of the nuclear wave packet, Eq. \eqref{dtrhoel_1}.
Accordingly, the magnitude of the perturbation to the electron flow introduced by the correlation length
in Eqs. \eqref{jc_left} and \eqref{taylor_jc_left} provides a natural criterion to evaluate a posteriori the quality of the approximation.
In the present example, the error introduced to the norm of the wavefunction by the BOBS ansatz remains below
$\sim1 \%$ at all times, well within the perturbative regime.

The introduction of a non-adiabatic perturbation proportional to the square of the correlation length, Eq. \eqref{taylor_jc_left}, does not alter the properties of the flow significantly.
On the other hand, the flux lines of the vector field obtained from the definition Eq. \eqref{taylor_jc_right} are obviously too localized around the nuclei.
Whereas the field lines are almost parallel to the nuclear motion in the non-Born-Oppenheimer simulations, the BOBS field show electrons flowing around
the cusps of the nuclear distribution. It was confirmed that the two vector fields are not related via a divergence-free gauge field. 
This behaviour of the correlated ansatz results from the mean-field character of the method, i.e., 
all three components of the electronic flux density are optimized using a single value of the correlation length $\delta_q$.
A better quantitative agreement with the non-Born-Oppenheimer flux density can be expected when treating separately the correlation length
for the electronic motions parallel and perpendicular to the nuclear vibration along $Q$.
This by no means reduces the validity of the qualitative picture discussed above, but care should be taken in over-analyzing the fields quantitatively, especially for the component perpendicular to the nuclear motion. 

\begin{figure}[bt]
\includegraphics{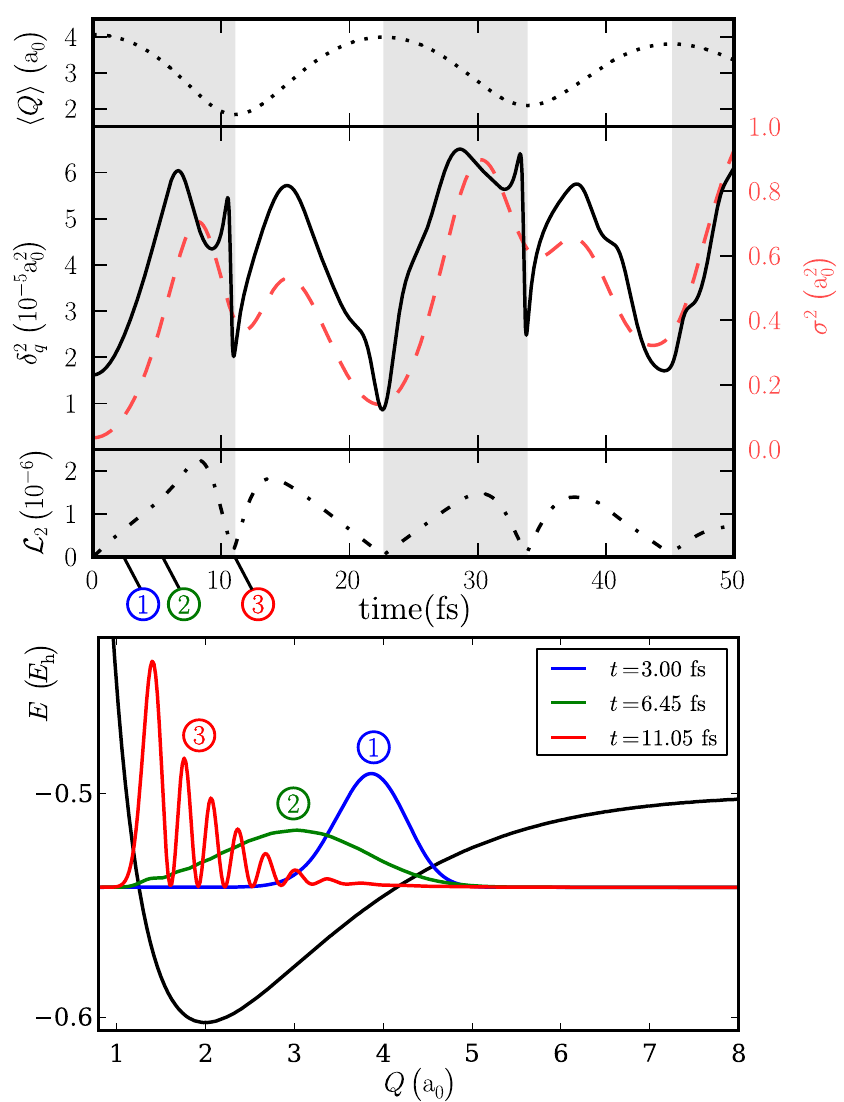}
\caption{\label{fig:dq_vs_variance} (Color online) Top panel: The expectation value of the internuclear distance $\left\langle Q\right\rangle$ (black dotted line) as a function of time.
Second panel: The squared value of the correlation length $\delta_q^2$ (black solid line) and 
the squared variance of the internuclear distance $\sigma^2 = \left\langle Q^2\right\rangle - \left\langle Q\right\rangle^2$ (red dashed line) as a function of time. 
Thrid panel: The error of the optimization per grid point, $\mathcal{L}_2=\frac{1}{N}\big\|\frac{\partial}{\partial t}\rho_{c}+\vec{\nabla}_{e}\cdot\vec{j}_{c}\big\|_{2}$, (black dashed dotted line) in units of $E_{\rm h}/(\hbar a_0^3)$ as a function of time. 
Regions, where the expectation value of the internuclear distance $\left\langle Q\right\rangle$ decreases, are shaded gray.
Notice the different scaling.
Bottom panel: Nuclear wavepacket at three times during the first half oscillation cycle: 
$t=3.00\,{\rm fs}$ (blue), $t=6.45\,{\rm fs}$ (green), and $t=11.05\,{\rm fs}$ (red).
}
\end{figure}

The second panel in Fig. \ref{fig:dq_vs_variance} depicts the time evolution of the correlation length squared, $\delta_q^2$, for two complete
nuclear oscillation periods. These can be identified from the expectation value of the internuclear distance, shown in the top panel.
A clear and unique optimal value of the correlation length is found at each timestep, which can further be seen to be very small and exhibit
a smooth periodic behavior. The dashed line in the central panel represents the squared variance of the nuclear wavepacket, $\sigma^2$.
Between the turning points, both the correlation length and the nuclear wavepacket variance behave similarly. This can be understood in simple physical terms:
as the nuclear wavepacket moves towards the bottom of the well, the nuclei acquire a larger velocity and the electrons drag on a longer scale behind the molecular
motion. This is captured by the correlation length, which weights the importance of non-adiabatic couplings and momentum transfer between nuclear and electronic
degrees of freedom.
At the turning points, the correlation length is subject to a more structured time evolution even though the nuclear variance remains smooth.

In order to confirm that the complex structure of $\delta_q^2$ conveys physical meaning, the residual error from the optimization in Eq. \eqref{opt},
$\mathcal{L}_2$, is reported per grid point in the third panel of Fig. \ref{fig:dq_vs_variance}. 
This error remains at least one order of magnitude smaller that the main features of the electron flow 
reported in Fig. \ref{fig:snapshots} and it correlates with the variance of the internuclear distance, $\sigma^2$.
Moreover, it is smoother and less structured than the time evolution of the correlation length.
This can be understood on the basis of the nuclear variance, as a wider spread of the underlying nuclear velocity field will
render minimization of Eq. \eqref{opt} with a single parameter less accurate.
Thus, the fast oscillating structures observed in the correlation length at the inner classical turning point - and to a lesser extent at the outer ones -
have a physical origin other than the nuclear dynamics.
\begin{figure}[tb]
\includegraphics{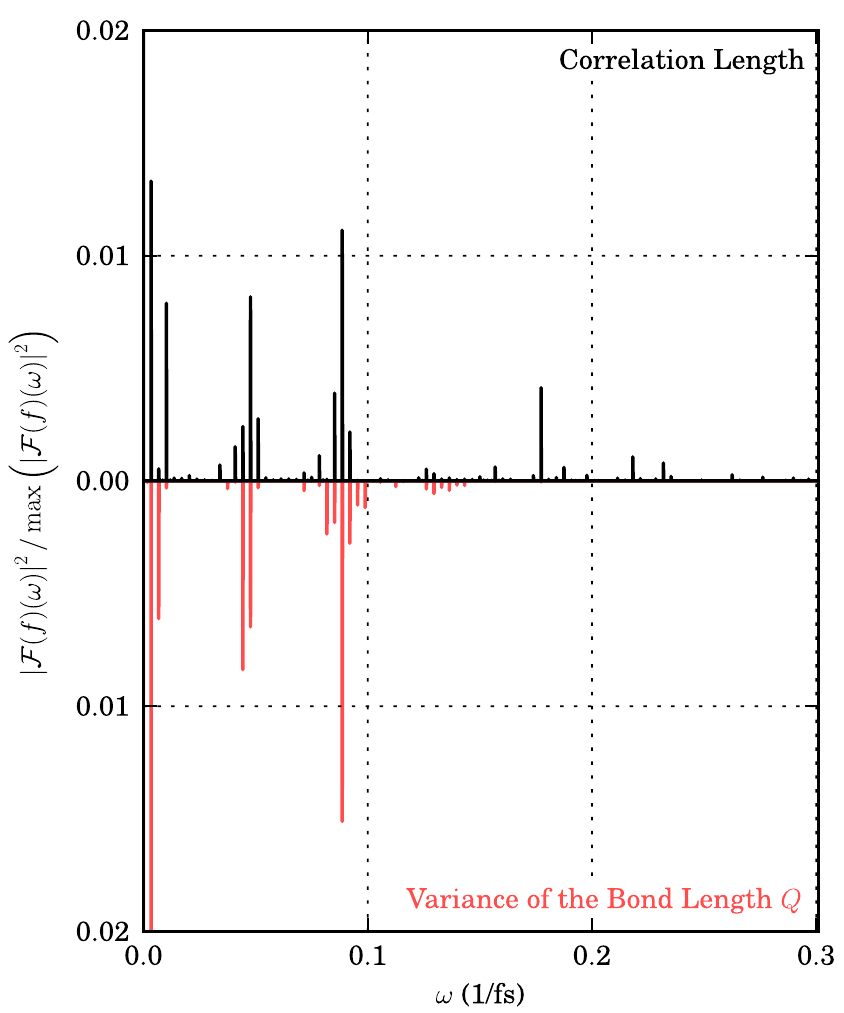}
\caption{\label{fig:fft_dq_vs_variance} (Color online) Normalized power spectra of the correlation length squared $\delta_q^2$ (black) and 
the variance of the internuclear distance $\sigma^2 = \langle Q^2\rangle-\langle Q\rangle^2$  (red) as a function of the frequency $\omega$. 
The dynamics was simulated up to twice the nuclear recurrence time of $T=293.16\,{\rm fs}$ with a time step width of $\Delta t=0.3\,{\rm fs}$.
}
\end{figure}

The comparison of the correlation length and the internuclear distance variance becomes clearer 
when looking at the power spectrum of their time evolution, as shown in Fig. \ref{fig:fft_dq_vs_variance}. 
Both quantities contain several common components at low frequencies, reflecting the classical behavior of the 
electrons instantaneously reacting to the nuclear motion between the turning points. 
However, the spectrum of $\delta_q^2$ is more complex and shows additional peaks at higher frequencies.
These are attributed to interference effects in the nuclear wavepacket upon reflection,
which can be observed in the bottom panel of Fig. \ref{fig:dq_vs_variance},
where the nuclear wavepackets at $t=3.00\,{\rm fs}$, $t=6.45\,{\rm fs}$, and $t=11.05\,{\rm fs}$ are depicted. The latter shows strongly
oscillatory structures, which explains the rapid variation of the correlation length at the 
turning points, as the electrons react more substantially to the intricate structure of the nuclear wavepacket.
As such, this is a purely quantum mechanical effect stemming from the implicit electron dynamics alone.

\section{Conclusions}

A fundamental problem of the Born-Oppenheimer approximation is that the electronic degrees of freedom are represented using real-valued wavefunctions,  
leading to the counterintuitive results that electrons remain stationary upon nuclear dynamics instead of flowing along with the molecular motion.
In this work, we presented a numerical approach to circumvent this intrinsic problem within the standard framework of the Born-Oppenheimer approximation: 
the Born-Oppenheimer Broken Symmetry (BOBS) approach to the adiabatic electronic flux density. 
In a very first step, we introduced a translation operator and applied it to the total (vibronic) density matrix operator. 
Tracing out the nuclear degrees of freedom, the electronic probability density could be forced to satisfy the electronic continuity equation approximately
using a numerical minimization procedure. This cost functional minimization depends on a single control parameter, 
the correlation length $\delta_q$, which we interpret as an electron-nucleus correlation in nuclear configuration space.
The electron flow and associated electronic flux density obtained from our Born-Oppenheimer Broken Symmetry procedure are real-valued,
as opposed to the AIMD-based ''time-shift`` flux approach of Okayama and Takatsuka.

The application of the translation operator to the electronic wavefunction in the position representation can become computationally prohibitively expensive,
and a Taylor series expansion of the correlated electron density to second order was introduced, yielding nearly identical results as the non-linear cost minimization.
The series expansion revealed that the physical origin of the electronic flux density observed in the BOBS ansatz is the first-order non-adiabatic coupling between
electrons and nuclei, with second order non-adiabatic contributions contributing to the electron flow.
Because of its computational efficiency, this approach could be easily applied to very large systems.

A vibrating hydrogen molecular ion ${\rm H}_2^+$ in the electronic ground state ${}^2\Sigma^+_g$ served as a test system for the BOBS approach. 
The electron flow was seen to be hardly affected by translation of the density matrix and compared well with the non-Born-Oppenheimer results, 
while the electronic flux density qualitatively recovered the correct features - symmetry at the inversion center and the turning point,
nodal planes perpendicular to the nuclear motion, etc - at all times.
Analysis of the time evolution of the correlation length demonstrated that, while a large portion of the implicit electron dynamics
can be correlated to the variance of the time-evolving nuclear wavepacket,
electrons show a stronger quantum mechanical character at the turning points due to interference effects upon reflection.

Because of its simplicity, the Born-Oppenheimer Broken Symmetry (BOBS) approach to the adiabatic electronic flux density 
appears as a possibly valuable semi-quantitative tool for understanding the electron dynamics of many other chemical processes.
A unique feature of the BOBS method is the explicit consistency enforcement of the electronic continuity equation at all times,
which simultaneously provides an error estimate.
In order to improve the quantitative predictive power of the method,
handling the different components of the electronic flux density with separate values for the correlation length
$\delta_q\to\{\delta_{q_x},\delta_{q_y},\delta_{q_z}\}$, 
could be used to reduce the undesirable localized behavior.

\begin{acknowledgments}
The authors gratefully acknowledge Jhon Fredy P\'erez Torres and Dennis J. Diestler for 
stimulating discussions. The authors thank the Scientific Computing
Services Unit of the Zentraleinrichtung f\"ur Datenverarbeitung
at Freie Universt\"at Berlin for allocation of computer time.
The funding of the Deutsche Forschungsgemeinschaft through the Emmy-Noether
program (project TR1109/2-1) and from the Elsa-Neumann foundation of the Land Berlin
is also acknowledged.
\end{acknowledgments}

\appendix
\section{Vibrational eigenstates representation}

For a given initial condition, the nuclear wavepacket $\chi$ can be expanded in the eigenstates basis of the associated nuclear Hamiltonian
$H_{\rm nuc}\chi_{n} = E_{n}\chi_{n}$ to
\label{sec:basis_set}
\begin{equation}\begin{aligned}
  \chi & =\sum_{n}a_{n}e^{-\imath E_{n}t/\hbar}\chi_{n}.
\end{aligned}\end{equation}
Accordingly, the derivatives of the nuclear wavepacket are defined as
\begin{equation}\begin{aligned} 
  \frac{\partial}{\partial t}\chi & =-\frac{\imath}{\hbar}\sum_{n}a_{n}E_{n}e^{-\imath E_{n}t/\hbar}\chi_{n} \\
  \chi'&=\nabla_{Q}\chi =\sum_{n}a_{n}e^{-\imath E_{n}t/\hbar}\chi_{n}'.
\end{aligned}\end{equation}
Inserting this basis set representation in the correlated continuity equation Eqs. (\ref{jc_ceq}-\ref{jc_right}) yields
\begin{equation}\begin{aligned}\label{basis_jc_left}
  \frac{\partial}{\partial t}&\rho_{c} =  -\int {\rm d}Q \left(\varphi^{-\delta_{q}}\varphi^{+\delta_{q}}\right)\\
  &\times\sum_{m<n}\omega_{mn}a_{n}a_{m}\left(\chi_{n}^{-\delta_{q}}\chi_{m}^{+\delta_{q}}+\chi_{n}^{+\delta_{q}}\chi_{m}^{-\delta_{q}}\right)\sin\left(\omega_{mn}t\right)\\
\end{aligned}\end{equation}
and
\begin{equation}\begin{aligned}\label{basis_jc_right}
  -\vec{\nabla}_{e}\cdot&\vec{j}_{c} = -\frac{\hbar}{m_{e}}\vec{\nabla}_{e}\cdot \int {\rm d}Q
  \left(\varphi^{+\delta_{q}}\vec{\nabla}_{e}\varphi^{-\delta_{q}}-\varphi^{-\delta_{q}}\vec{\nabla}_{e}\varphi^{+\delta_{q}}\right) \\
  &\times\sum_{m<n}a_{n}a_{m}\left(\chi_{n}^{-\delta_{q}}\chi_{m}^{+\delta_{q}}-\chi_{n}^{+\delta_{q}}\chi_{m}^{-\delta_{q}}\right)\sin\left(\omega_{mn}t\right)
\end{aligned}\end{equation}
with $\omega_{mn}=\frac{E_{m}-E_{n}}{\hbar}$ and $a_{n}\in\Re$. 
Eqs. (\ref{basis_jc_left}, \ref{basis_jc_right}) can be trivially extended to complex-valued expansion coefficients $a_n$.
Note that the left [cf. Eq. (\ref{basis_jc_left})] and the right-hand-side [cf. Eq. (\ref{basis_jc_right})]
of the correlated continuity equation in the basis set representation oscillate with the same sinusoidal behavior.
After some manipulations, the Taylor series expansion Eqs. (\ref{taylor_jc_left}, \ref{taylor_jc_right}) thus yields
\begin{widetext}
  \begin{eqnarray}
    \frac{\partial}{\partial t}\rho_{c} &=& -2\int {\rm d}Q \left(\left|\varphi\right|^{2} 
    +\delta_{q}^{2}\left(\varphi''\varphi-\varphi'\varphi'\right)\right)\sum_{m<n}\omega_{mn}a_{n}a_{m}\chi_{n}\chi_{m}\sin\left(\omega_{mn}t\right) \nonumber\\
    & &-\delta_{q}^{2}\int {\rm d}Q \left|\varphi\right|^{2}\sum_{m<n}\omega_{mn}a_{n}a_{m}\left(\chi_{n}''\chi_{m}-2\chi_{n}'\chi_{m}'+\chi_{n}\chi_{m}''\right)\sin\left(\omega_{mn}t\right) 
    +O\left(\delta_{q}^{4}\right)\\
    -\vec{\nabla}_{e}\cdot\vec{j}_{c} &=& +\frac{2\hbar\delta_{q}^{2}}{m_{e}}\vec{\nabla}_{e}\cdot \int {\rm d}Q \left(\varphi\vec{\nabla}_{e}\varphi'-\varphi'\vec{\nabla}_{e}\varphi\right) 
    \sum_{m<n}a_{n}a_{m}\left(\chi_{n}\chi_{m}'-\chi_{n}'\chi_{m}\right)\sin\left(\omega_{mn}t\right)+O\left(\delta_{q}^{4}\right).
  \end{eqnarray}
\end{widetext}

\bibliography{correlated_efd}

\end{document}